\documentclass[10pt, journal]{IEEEtran}
\usepackage{graphicx,subfigure}
\usepackage{cite}
\usepackage{amssymb}
\usepackage{bm}
\usepackage{authblk}

\author{Abdeljalil Beniiche\thanks{abdeljalil.beniiche@emt.inrs.ca}}

\affil{INRS, Montr\'eal, QC, Canada}

\title{A Study of Blockchain Oracles}

\begin{document}

\date
\tableofcontents
\newpage
\maketitle 

\begin{abstract}

The limitation with smart contracts is that they cannot access external data which might be required to control the execution of business logic. \emph{Oracles} can be used to provide external data to smart contracts. An oracle is an interface that delivers data from external data outside the blockchain to a smart contract to consume. Oracle can deliver different types of data depending on the industry and requirements. In this paper, we study and describe the widely used blockchain oracles. Then, we elaborate on his potential role, technical architecture, and design patterns. Finally, we discuss the human oracle and his key role in solving the truth problem by reaching a consensus about a certain inquiry and tasks.

\end{abstract}

\section{Introducing Blockchain Oracle}

Blockchain oracles are third-party services that provide smart contracts with external information. They serve as bridges between blockchains and the outside world. Blockchains and smart contracts cannot access off-chain data (data that is outside of the network). However, for many contractual agreements, it is vital to have relevant information from the outside world to execute the agreement.

This is where blockchain oracles come into play, as they provide a link between off-chain and on-chain data. Oracles are vital within the blockchain ecosystem because they broaden the scope in which smart contracts can operate. Without blockchain oracles, smart contracts would have very limited use as they would only have access to data from within their networks. 

It's important to note that a blockchain oracle is not the data source itself, but rather the layer that queries, verifies, and authenticates external data sources and then relays that information. The data transmitted by oracles comes in many forms, e.g., information, the successful completion of a payment, or the temperature measured by a sensor. 

To call data from the outside world, the smart contract has to be invoked, and network resources have to be spent. Some oracles also have the ability to not only relay information to smart contracts but to send it back to external sources.

Many different types of oracles exist with different functions or characteristics, how a blockchain oracle operates is entirely dependent on what it is designed for. After briefly describing all types of blockchain oracles, this paper will go through some of those designs, we will study and describe the widely used blockchain oracles. Then, we elaborate on his potential role, technical architecture, and design patterns. Then, we discuss the human oracle and its role to solving the truth problem by reaching a consensus about a certain inquiry. Finally, we give some examples of oracles use cases and we conclude the paper.

\section{Types of Blockchain Oracle}

Blockchain oracles can be classified depending on a number of different qualities: 

\begin{itemize}
\item \emph{Source}: does the data originate from software, hardware, or human?
\end{itemize}

\begin{itemize}
\item \emph{Direction of information}: is it inbound or outbound?
\end{itemize}

\begin{itemize}
\item \emph{Trust:} is it centralized or decentralized?
\end{itemize}

A single oracle can fall into multiple categories. For example, an oracle that sources information from a company website is a centralized inbound software oracle.

\subsection{Software Oracles}

Software oracles (also known as deterministic oracles) interact with online sources of information and transmit it to the blockchain. This information can come from online databases, servers, websites, essentially, any data source on the Web.

The fact that software oracles are connected to the Internet not only allows them to supply information to smart contracts but also to transmit that information in real-time. This makes them one of the most common types of blockchain oracles. 

Information typically provided by software oracles can include exchange rates, digital asset prices, real-time flight information, or any other information we need it.

\subsection{Hardware  Oracles}

Some smart contracts need to interface with the real world. Hardware oracles are designed to get information from the physical world and make it available to smart contracts. Such information could be relayed from electronic sensors, IoT, barcode/QR scanners, RFID tags, robot, and other information reading devices.

A hardware oracle essentially ``translates" real-world events into digital values that can be understood by smart contracts.

An example of this could be a sensor that checks if a truck transporting goods has arrived at a loading bay. If it does, it relays the information to a smart contract that can then execute decisions based on it. A concrete example of hardware oracle is implemented in Supply Chain.

\subsection{Human Oracles}

Sometimes individuals with specialized knowledge/skills in a particular field can also serve as oracles. They can research and verify the authenticity of information from various sources and translate that information to smart contracts. Since human oracles can verify their identity using cryptography, the possibility of a fraudster faking their identity and providing corrupted data is relatively very low.

Human oracles are not only able to transmit deterministic data, but also to respond to arbitrary inquiries, which could potentially be hard to do by a machine. Answers to inquiries for unstructured data, like  ``did candidate X win the Canadian election?" or  ``did sports team A beat sports team B?" are somewhat complex to be automatically deducted by a computer program. Hence, Human oracles which respond to arbitrary inquiries and provide manual input are imperative to certain smart contract and decentralized applications.

In Sec. 4, we will discuss in greater detail the role of the human in the context of blockchain oracles, citing some existing examples.

\subsection{Computation Oracles}

So far, we have only discussed oracles in the context of requesting and delivering data (also knows as \emph{Data Carrier Oracles} or \emph{Automated Oracles}). However, oracles can also be used to perform arbitrary ``off-chain" computation solution, a function that can be especially useful given Ethereum's inherent block gas limit and comparatively expensive computation cost. 

Rather than just relaying the results of a query, computation oracles can be used to perform computation on a set of inputs and return a calculated result that may have been infeasible to calculate on-chain. For example, one might use a computation oracle to perform a computationally intensive regression calculation in order to estimate the yield of a bond contract.

\subsection{Inbound/Outbound Oracles}

Inbound oracles transmit information from external sources to smart contracts, while outbound oracles send information from smart contracts to the external world.

An example of an inbound oracle is one that tells a smart contract what the temperature is measured by a sensor. An example of an outbound oracle can be considered with a smart lock. If funds are deposited to an address, the smart contract sends this information through an outbound oracle to a mechanism that unlocks the smart lock.

\subsection{Contract-specific Oracles}

A contract-specific oracle is one that is designed to be used by a single smart contract. This means that if one wants to deploy several smart contracts, a proportionate number of contract-specific oracles have to be developed.

This type of oracle is considered very time-consuming and expensive to maintain. Companies that want to extract data from a variety of sources may find this approach very impractical. On the other hand, since contract-specific oracles can be designed from scratch to serve a specific use case, developers have high flexibility to tailor them to specific requirements.

\subsection{Consensus-based Oracles}

In contrast to software oracles, consensus-based oracles do not use a single source. There are several ways to create and use decentralized oracles. One would be a rating system inside a prediction market. To reduce risk and provide more security, a combination of oracles might be used. For example you could take the average of 5 oracles. Or, 5 out of 7 oracles can determine the outcome of an event. 

Consensus-based oracles are slower, only because it takes more time to reach consensus. If we can not trust an oracle that gives us the  information we need, or if we are not certain of the accuracy at all times, a slightly safer path may be through consensus-based oracles. Especially if there exists a large sum of money or legal ramifications behind it.

\section{Centralized/Decentralized Oracles}

A centralized oracle is controlled by a single entity and is the sole provider of information for the smart contract. Using only one source of information can be risky, the effectiveness of the contract depends entirely on the entity controlling the oracle. Also, any malicious interference from a bad actor will have a direct impact on the smart contract. The main problem with centralized oracles is the existence of a single point of failure, which makes the contracts less resilient to vulnerabilities and attacks.

Decentralized oracles share some of the same objectives as the public blockchains and avoiding counter-party risk. They increase the reliability of the information provided to smart contracts by not relying on a single source of truth. The smart contract queries multiple oracles to determine the validity and accuracy of the data, this is why decentralized oracles can also be referred to as consensus oracles.

Some blockchain projects provide decentralized oracle services to other blockchains. Decentralized oracles can also be useful in prediction markets, where the validity of a certain outcome can be verified by social consensus.

While decentralized oracles aim to achieve trustessness, it is important to note that just like trust-less blockchain networks, decentralized oracles do not completely eliminate trust, but rather distribute it between many participants.

\subsection{Oraclize (now is ``Provable Things")}

\subsubsection{General Concepts}

Now, let's talk about the most widely used oracle in the blockchain era, \emph{Oraclize} or \emph{Provable Things}. Provable is the leading oracle service for smart contracts and blockchain applications, serving thousands of requests every day on platforms like Ethereum, R3 Corda, Hyperledger Fabric and EOS.

As explained before, in the blockchain space, an oracle is a party which provides data. The need for such figure arise from the fact that blockchain applications, such as smart contracts cannot access and fetch directly the data they require: price feeds for assets and financial applications; weather-related information for peer-to-peer insurance; random number generation for gambling.

But to rely on a new trusted intermediary, the oracle in this case, it would be betraying the security and reduced-trust model of blockchain applications: which is what makes them interesting and useful in first place.

One solution is to accept data inputs from more than one untrusted or partially trusted party and then execute the data-dependent action only after a number of them have provided the same answer or an answer within some constrains. This type of system can be considered a decentralized oracle system. Unfortunately, this approach has severe limitations:

\begin{itemize}
\item It requires a predefined standard on data format.
\end{itemize}

\begin{itemize}
\item It is inherently inefficient: all the parties participating will require a fee and, for every request, it will take time before reaching a sufficient number of answers.
\end{itemize}

The solution developed by Provable (Oraclize) is instead to demonstrate that the data fetched from the original data-source is genuine and untampered. This is accomplished by accompanying the returned data together with a document called \emph{authenticity proof}. The authenticity proofs can build upon different technologies such as auditable virtual machines and Trusted Execution Environments.

This solution elegantly solves the oracle problem:

\begin{itemize}
\item Blockchain Application's developers and the users of such applications don't have to trust Provable; the security model is maintained.
\end{itemize}

\begin{itemize}
\item Data providers don't have to modify their services in order to be compatible with blockchain protocols. Smart contracts can directly access data from Web sites or APIs.
\end{itemize}

\begin{itemize}
\item Provable engine can be easily integrated with both private and public instances of different blockchain protocols.
\end{itemize}

While building the service, the Provable team has realized that the concept of authenticity proofs has much broader applicability than initially envisioned. For example, the Provable Random Data-source can be used even by traditional gambling applications to ensure users of continuous fairness of operation

\subsubsection{Provable Engine}

The Provable Engine powers the service for both blockchain-based and non-blockchain-based application. Internally replicates an ``If This Then That" logical model. This means that it will execute a given set of instructions if some other given conditions are met. For example, it could repeatedly verify a condition and only return data or perform an action when the condition has been met. This flexibility enables the engine to be leveraged in many different ways and contexts, even outside of the blockchain context.

A valid request for data to Provable, done via the native blockchain integration or via the HTTP API, should specify the following arguments:

\begin{itemize}
\item A data source type
\end{itemize}

\begin{itemize}
\item A query
\end{itemize}

\begin{itemize}
\item Optionally, an authenticity proof type
\end{itemize}

\subsubsection{Data Source Types}

A data source is a trusted provider of data. It can be a website or web API such as Reuters, Weather.com, BBC.com, or a secure application running on an hardware-enforced Trusted Execution Environment (TEE) or an auditable, locked-down virtual machine instance running in a cloud provider. Provable currently offers the following types of native data sources:

\begin{itemize}
\item URL: enables the access to any webpage or HTTP API endpoint.
\end{itemize}

\begin{itemize}
\item WolframAlpha\footnote{WolframAlpha is a computational knowledge engine or answer engine. It is an online service that answers factual queries directly by computing the answer from externally sourced ``curated data", rather than providing a list of documents or web pages that might contain the answer as a search engine might.}: enables native access to WolframAlpha computational engine.
\end{itemize}

\begin{itemize}
\item IPFS: provides access to any content stored on an IPFS file.
\end{itemize}

\begin{itemize}
\item Random: provides untampered random bytes coming from a secure application running on a hardware wallet such as Ledger Nano S.
\end{itemize}

\begin{itemize}
\item Computation: provides the result of arbitrary computation.
\end{itemize}

Additionally, there also some meta data source such as:

\begin{itemize}
\item nested: enables the combination of different types of data source or multiple requests using the same data source, and it returns a unique result.
\end{itemize}

\begin{itemize}
\item identity: it returns the query.
\end{itemize}

\begin{itemize}
\item decrypt: it decrypts a string encrypted to the Provable private key.
\end{itemize}

\subsubsection{Query}

A query is an array of parameters which needs to be evaluated in order to complete a specific data source type request: ``query: ( parameter1, parameters2, ...)".

The first parameter is the main argument and it is usually mandatory. For example, in the case of the URL Data Source Type, the first argument is the expected URL where the resource resides. If only the first argument is present, then the URL Data Source assumes that an HTTP GET was requested. The second parameters, which it is optional, should contain the data payload of the HTTP POST request.

The intermediate result of a query may need to be parsed: for example, to extract a precise field in JSON API response. Therefore, the query can also specify parsing helpers to be applied.

\subsubsection{Parsing Helpers}

Provable offers JSON, XML, XHTML and a binary parser helpers. For examples:

\textbf{JSON Parsing:} To extract the value of a specific element from a JSON document, we may use our built-in JSON parser. An example usage case, for extracting the ETH/USD price field from the Kraken API, serving it as a JSON doc, is by surrounding the API endpoint in question with the helper.

\textbf{XML Parser:} To extract the value of a specific element from an XML document, we may use our built-in XML parser. An example usage case, for extracting the diesel price from an API serving it as an XML doc, is by surrounding the API endpoint in question with the helper.

\textbf{HTML Parser:} Useful for HTML scraping. The desired XPATH can be specified as argument of xpath(..).

\textbf{Binary Helper:} It can be useful to extract parts of a binary intermediate result by using the slice (offset, length) operator. The first parameter is the expected to be the offset, while the second one is the length of the returned slice. 

\subsubsection{Authenticity Proofs}

Authenticity proofs are at the core of Provable's oracle model. Provable or Oraclize provides a secure connection between smart contracts and the external world, enabling both data-fetching and delegation of code execution. The data (or result) is delivered to the smart contract along with a so-called ``authenticity proof", a cryptographic guarantee proving that such data (or result) was not tampered with. By verifying the validity of such authenticity proof, anybody at any time can verify whether the data (or result) delivered is authentic or not.

\subsection{ChainLink}
\subsubsection{Overview}

While centralized oracles suffice for many applications, they represent single points of failure in the Ethereum network. A number of schemes have been proposed around the idea of decentralized oracles as a means of ensuring data availability and the creation of a network of individual data providers with an on-chain data aggregation system.

ChainLink's core functional objective is to bridge two environments: on-chain and off-chain. ChainLink will initially be built on Ethereum, but team intends for it to support all leading smart contract networks for both off-chain and cross-chain interactions. In both its on and off-chain versions, ChainLink has been designed with modularity in mind. Every piece of the ChainLink system is upgradable, so that different components can be replaced as better techniques and competing implementations arise.

\subsubsection{On-Chain Architecture}

As an oracle service, ChainLink nodes return replies to data requests or queries made by or on behalf of a user contract, which ChainLink's team refers to as requesting contracts and denote by USER-SC. ChainLink's on-chain interface to requesting contracts is itself an on-chain contract that ChainLink's team denotes by CHAINLINK-SC.

Behind CHAINLINK-SC, ChainLink has an on-chain component consisting of three main contracts:

\begin{itemize}
\item Reputation contract,
\end{itemize}

\begin{itemize}
\item Order-matching contract,
\end{itemize}

\begin{itemize}
\item Aggregating contract.
\end{itemize}

The reputation contract keeps track of oracle-service-provider performance metrics. The order-matching smart contract takes a proposed service level agreement (SLA), logs the SLA parameters, and collects bids from oracle providers. It then selects bids using the reputation contract and finalizes the oracle SLA. The aggregating contract collects the oracle providers' responses and calculates the final collective result of the ChainLink query. It also feeds oracle provider metrics back into the reputation contract. ChainLink contracts are designed in a modular manner, allowing for them to be configured or replaced by users as needed.

The on-chain work flow has three steps:

\begin{enumerate}
\item Oracle selection,
\item Data reporting,
\item Result aggregation.
\end{enumerate}

\textbf{Oracle Selection.} An oracle services purchaser specifies requirements that make up a SLA proposal. The SLA proposal includes details such as query parameters and the number of oracles needed by the purchaser. Additionally, the purchaser specifies the reputation and aggregating contracts to be used for the rest of the agreement.

Using the reputation maintained on-chain, along with a more robust set of data gathered from logs of past contracts, purchasers can manually sort, filter, and select oracles via off-chain listing services. ChainLink's team intentions is for ChainLink to maintain one such listing service, collecting all ChainLink-related logs and verifying the binaries of listed oracle contracts. The data used to generate listings will be pulled from the blockchain, allowing for alternative oracle-listing services to be built. Purchasers will submit SLA proposals to oracles off-chain, and come to agreement before finalizing the SLA on-chain.

Manual matching is not possible for all situations. For example, a contract may need to request oracle services dynamically in response to its load. Automated solutions solve this problem and enhance usability. For these reasons, automated oracle matching is also being proposed by ChainLink through the use of order-matching contracts.

Once the purchaser has specified their SLA proposal, instead of contacting the oracles directly, they will submit the SLA to an order-matching contract. The submission of the proposal to the order-matching contract triggers a log that oracle providers can monitor and filter based on their capabilities and service objectives. ChainLink nodes then choose whether to bid on the proposal or not, with the contract only accepting bids from nodes that meet the SLA's requirements. When an oracle service provider bids on a contract, they commit to it, specifically by attaching the penalty amount that would be lost due to their misbehavior, as defined in the SLA.
Bids are accepted for the entirety of the bidding window. Once the SLA has received enough qualified bids and the bidding window has ended, the requested number of oracles is selected from the pool of bids. Penalty payments that were offered during the bidding process are returned to oracles who were not selected, and a finalized SLA record is created. When the finalized SLA is recorded it triggers a log notifying the selected oracles. The oracles then perform the assignment detailed by the SLA.

\textbf{Data Reporting.} Once the new oracle record has been created, the off-chain oracles execute the agreement and report back on-chain.

\textbf{Result Aggregation.} Once the oracles have revealed their results to the oracle contract, their results will be fed to the aggregating contract. The aggregating contract tallies the collective results and calculates a weighted answer. The validity of each oracle response is then reported to the reputation contract. Finally, the weighted answer is returned to the specified contract function in USER-SC.

Detecting outlying or incorrect values is a problem that is specific to each type of data feed and application. For instance, detecting and rejecting outlying answers before averaging may be necessary for numeric data but not boolean. For this reason, there will not be a specific aggregating contract, but a configurable contract address which is specified by the purchaser. ChainLink will include a standard set of aggregating contracts, but customized contracts may also be specified, provided they conform to the standard calculation interface.

\subsubsection{Off-Chain Architecture}

Off-chain, ChainLink initially consists of a network of oracle nodes connected to the Ethereum network, and ChainLink's team intends for it to support all leading smart contract networks. These nodes independently harvest responses to off-chain requests. Their individual responses are aggregated via one of several possible consensus mechanisms into a global response that is returned to a requesting contract USER-SC. The ChainLink nodes are powered by the standard open source core implementation which handles standard blockchain interactions, scheduling, and connecting with common external resources. Node operators may choose to add software extensions, known as external adapters, that allow the operators to offer additional specialized off-chain services. ChainLink nodes have already been deployed alongside both public blockchains and private networks in enterprise settings; enabling the nodes to run in a decentralized manner is the motivation for the ChainLink network.

\textbf{ChainLink Core.} The core node software is responsible for interfacing with the blockchain, scheduling, and balancing work across its various external services. Work done by ChainLink nodes is formatted as assignments. Each assignment is a set of smaller job specifications, known as subtasks, which are processed as a pipeline. Each subtask has a specific operation it performs, before passing its result onto the next subtask, and ultimately reaching a final result. ChainLink's node software comes with a few subtasks built in, including HTTP requests, JSON parsing, and conversion to various blockchain formats.

\textbf{External Adapters.} Beyond the built-in subtask types, custom subtasks can be defined by creating adapters. Adapters are external services with a minimal REST API. By modeling adapters in a service-oriented manner, programs in any programming language can be easily implemented simply by adding a small intermediate API in front of the program. Similarly, interacting with complicated multi-step APIs can be simplified to individual subtasks with parameters.

\textbf{Subtask Schemas.} The ChainLink team anticipates that many adapters will be open sourced, so that services can be audited and run by various community members. With many different types of adapters being developed by many different developers, ensuring compatibility between adapters is essential.

ChainLink currently operates with a schema system based on JSON Schema, to specify what inputs each adapter needs and how they should be formatted. Similarly, adapters specify an output schema to describe the format of each subtask's output.

\subsubsection{ChainLink Workflow}

Fig. 1 describe the Chainlink workflow. We note that, a requests could be distributed across both oracles and data sources. Fig. 2 shows an example of such two-level distribution.

\begin{figure*}[hbt!]
        \centering  
        \includegraphics[width=13cm]{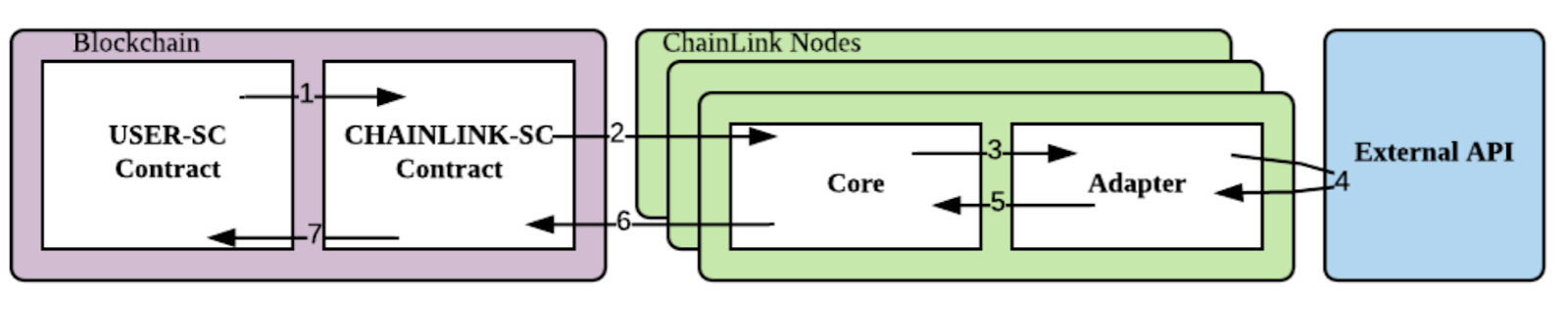}
        \caption{1) USER-SC makes an on-chain request; 2) CHAINLINK-SC logs an event for the oracles; 3) ChainLink core picks up the event and routes the assignment to an adapter; 4) ChainLink adapter performs a request to an external API; 5) ChainLink adapter processes the response and passes it back to the core; 6) ChainLink core reports the data to CHAINLINK-SC; 7) CHAINLINK-SC aggregates responses and passes them back as a single response to USER-SC (\emph{Source: https://chain.link}).}
\label{fig1}
\end{figure*}

\begin{figure*}[hbt!]
        \centering  
        \includegraphics[width=12cm]{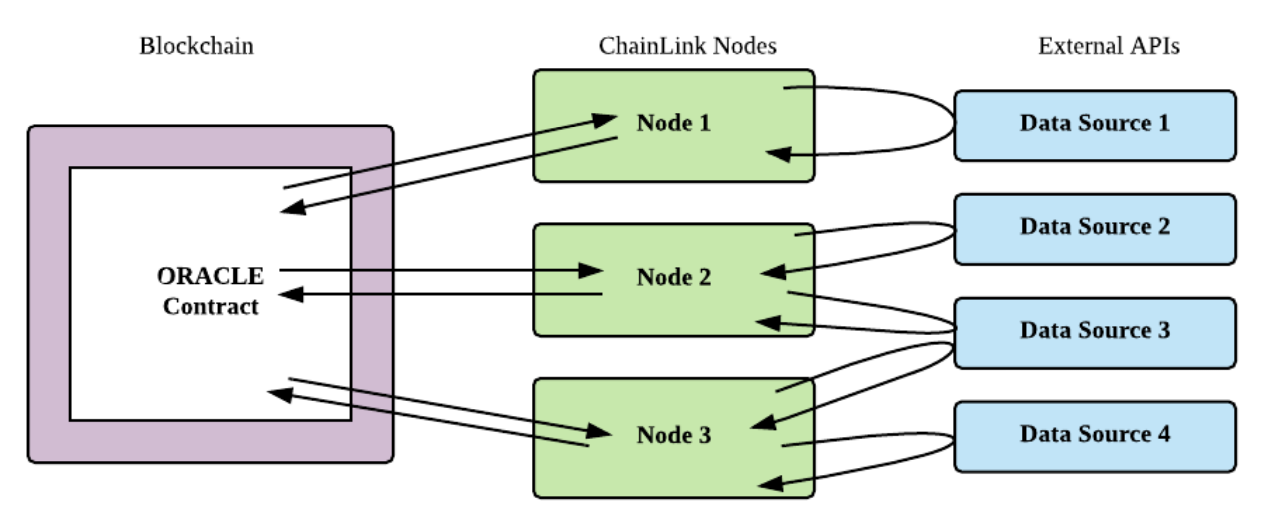}
        \caption{Requests are distributed across both oracles and data sources (\emph{Source: https://chain.link}).}
\label{fig1}
\end{figure*}

\subsubsection{Commit-Reveal Scheme}

A commitment scheme is a cryptographic algorithm used to allow someone to commit to a value while keeping it hidden from others with the ability to reveal it later. The values in a commitment scheme are binding, meaning that no one can change them once committed. The scheme has two phases: 

\begin{enumerate}
\item The commit phase: where a value is chosen and specified.
\item The reveal phase: where the chosen value is revealed and verified.
\end{enumerate}

To get a better understanding, consider this simplified example. Imagine Alice, the sender, placing a message in a locked box and handing it to Bob, the receiver. Bob can't access the message because it's locked in the box, and Alice can't change the message because it's in Bob's possession, but when Alice wants to reveal the message, she can unlock the box and show the message to Bob.

\section{Human Oracles}

When an automated oracle (software, hardware) can be pictured as a delivery guy, picking up a piece of information and delivering it to the customer which is the blockchain, human oracles do not only deliver but also manufacture this package. In other words, human oracles ``create" the truth by reaching a consensus about a certain inquiry. Because of this, human oracles can also be considered multi-source oracles. After explaining the oracle background of this ``truth by consensus" and reviewing its present use at prediction markets, it will be examined how this technique can be used for structured data as well.

\subsection{Background}

The general idea behind creating truth by consensus is to present several human individuals with a questionnaire or reporting card and give them an incentive to respond truthfully. This incentive would usually consist of some kind of tradeable asset with a monetary value such as a token. Vitalik Buterin was the first one to conceptualize a trusted data feed that utilizes aspects of behavioral economics with the proposal of SchellingCoin\footnote{V. Buterin, ``SchellingCoin: A Minimal-Trust Universal Data Feed," 2014}. This proposal built upon the game-theoretical research of Thomas Schelling.

Schelling elucidated the behavior of individuals within a relationship that does not share a transparent exchange of information. He introduced the thought that, even in the absence of communication, ``every situation provides some clue for coordinating behavior, some focal point of expectation of what the other expects him to expect to be expected to do." In other words, even without prior interaction, people are usually able to anticipate other participants' actions to some degree.

In the context of oracles, the behavioral pattern can be harnessed in order to agree on a certain state of fact. Assuming that the majority of people will act upon the motive of personal profit, they will provide information of which they think most people deem correct. The compulsion to place a deposit and the possibility to challenge reporting results can support this motivation to act truthfully and impede Sybil attacks at the same time. Due to this, collusion will become close to impossible provided there is a sufficient size and distribution of the network.

This however involves a considerable caveat: it is important to remember that Schelling's approach does not promote a single truth, but rather an attempt to predict human behavior. As data implemented to the blockchain is consistently regarded as true by smart contracts, truth is rather not transferred, but created by the oracle itself. This Schelling Point-based perception of truth might however not be concordant with the facts after all. Even if the majority of people consider something true, it does not necessarily mean it really is. On the blockchain however, this is indeed the case since informational input cannot be corrected or altered retroactively. While the deeper meaning of ``truth" is of a philosophical nature and cannot sufficiently be discussed at this point, it can be agreed that the nature of human perception is influenced by gullibility and vulnerability to wishful thinking. Hence, such ``truth by consensus" is to be treated with caution, to say the least. 

Also, any factor that was not determined may result in the inquiry falling apart. To give an example; as the seemingly clear question ``What is the price of Bitcoin at the 1st of January 2018 at 00:01?" does not involve a time zone, some people may assume it is EST + 0 and some will assess upon the basis of the inquirer's residence, while others might not even consider another time zone than their own.

Additionally, the same reservations as to conducting multiple data sources apply: there might be a common misinformation about a certain fact. Even if all the participants undertake to perform honestly, they might agree on something they are collectively misinformed about.

\subsection{Application in Prediction Markets}
Despite the abovementioned reservations, the approach of using human oracles in order to deduce a Schelling Point-based truth is the preferred method when it comes to decentralized prediction markets: both Augur and Gnosis (currently the largest decentralized prediction markets) do in some way include human oracles that are based on reaching a consensus of facts.

Augur\footnote{https://www.augur.net/} provides a solution in which the participants hold ``Reputation Tokens" which obligate them to report on a number of inquired events on a regular basis. These tokens can be split and traded just like cryptocurrencies. Reporting is only allowed until a preset quorum is met. The calculation of the result is based on the consensus algorithm proposed by Truthcoin\footnote{P. Sztorc,``Peer-to-Peer Oracle System and Prediction Marketplace," 2015.}. Through this act of reporting, reputation holders provide a trusted data feed for the prediction market. Other examples for this approach are Aeternity and Reality Token.

Gnosis\footnote{https://gnosis.io/} approach is different, but ultimatly resorts to human oracles as well: mainly, Gnosis derives information from centralized oracle services (using RealityKeys and Oraclize), but enables the users to challenge those results, creating something they refer to as ``The Ultimate Oracle". When challenging the provided results, the individual has to raise an amount of 100 Ether which functions as a deposit. Any Ethereum holder can now either support this challenge or dispute it by raising Ether. After a preset ammount of time the market is resolved and the ``winning" parties will distribute the deposits of the counterparty as ``earnings", additionally to the payout conducted by the market. This approach is prone to failure for obvious reasons: 100 Ether are currently worth aboout 33.138 USD or 28.091 EUR. This leads to a situation where a challanging party with enough resources could easily bully other participants into unjustifiedly losing their money. While this could theoretically be prevented by the community, most Ethereum holders will hardly care about the results of some random Gnsosis market.

\subsection{Adoption to Structured Data}

As of yet, human oracles are predominantly being used to address arbitrary inquiries. However, the approach could easily be applied to structured data as well with integrating a consensus as a ``filter" between the structured data and the result. This would require a manual input of structured data to be conducted by humans. As the immutability of the system grows with the size of its participants, a large amount of people would have to manually compare and type in structured data for best results. However, the practical transposition of this is presumably unlikely, as the ordinary market participant has neither the time nor the inclination to do so. But even if there was an occupation such as ``full-time oracle", human oracles do have intrinsic disadvantages: ``as human cognition is costly and slow, manual-input oracles are resource-intensive, not real time, and can handle only a limited set of questions at any given time".

The proposition of creating ``Reporting Pools" has been made as a refutation to these concerns. However, this would increase the possibility for a single malevolent actor to influence the oracles' voting-process in a fraudulent manner. Therefore, voting pools could constitute a potential point of attack and thus reintroduce a single point of failure. Hence, voting pools do not present a comprehensive solution to this problem. Therefore, while human oracles might introduce a trust less data feed, a scenario in which they completely replace automated software or hardware oracles seems unlikely. The challenges connected with the oracle problem arise with the reintroduction of a single point of failure, which leads the necessity of trusting the inquired data source and oracle. Hence, they can only be sufficiently resolved by eliminating this possibility for oracles and data sources to influence the process of data collection and implementation.

Fig. 3 and Fig. 4 depicts the automated oracle (hardware, software) and the human oracle structures.

\begin{figure*}[h]
        \centering  
        \includegraphics[width=12cm]{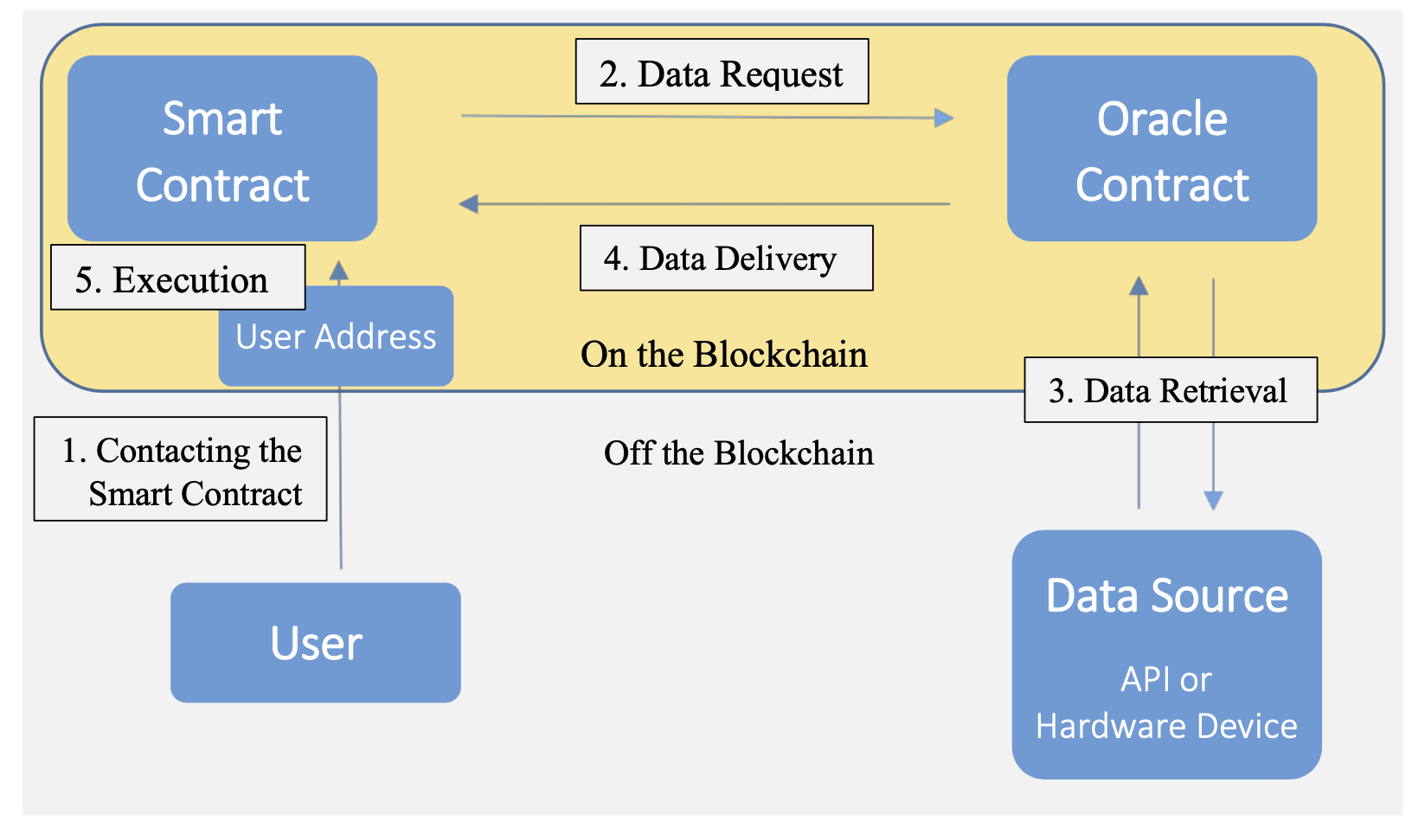}
        \caption{Automated  oracle structure}
\label{fig1}
\end{figure*}

\begin{figure*}[hbt!]
        \centering  
        \includegraphics[width=12cm]{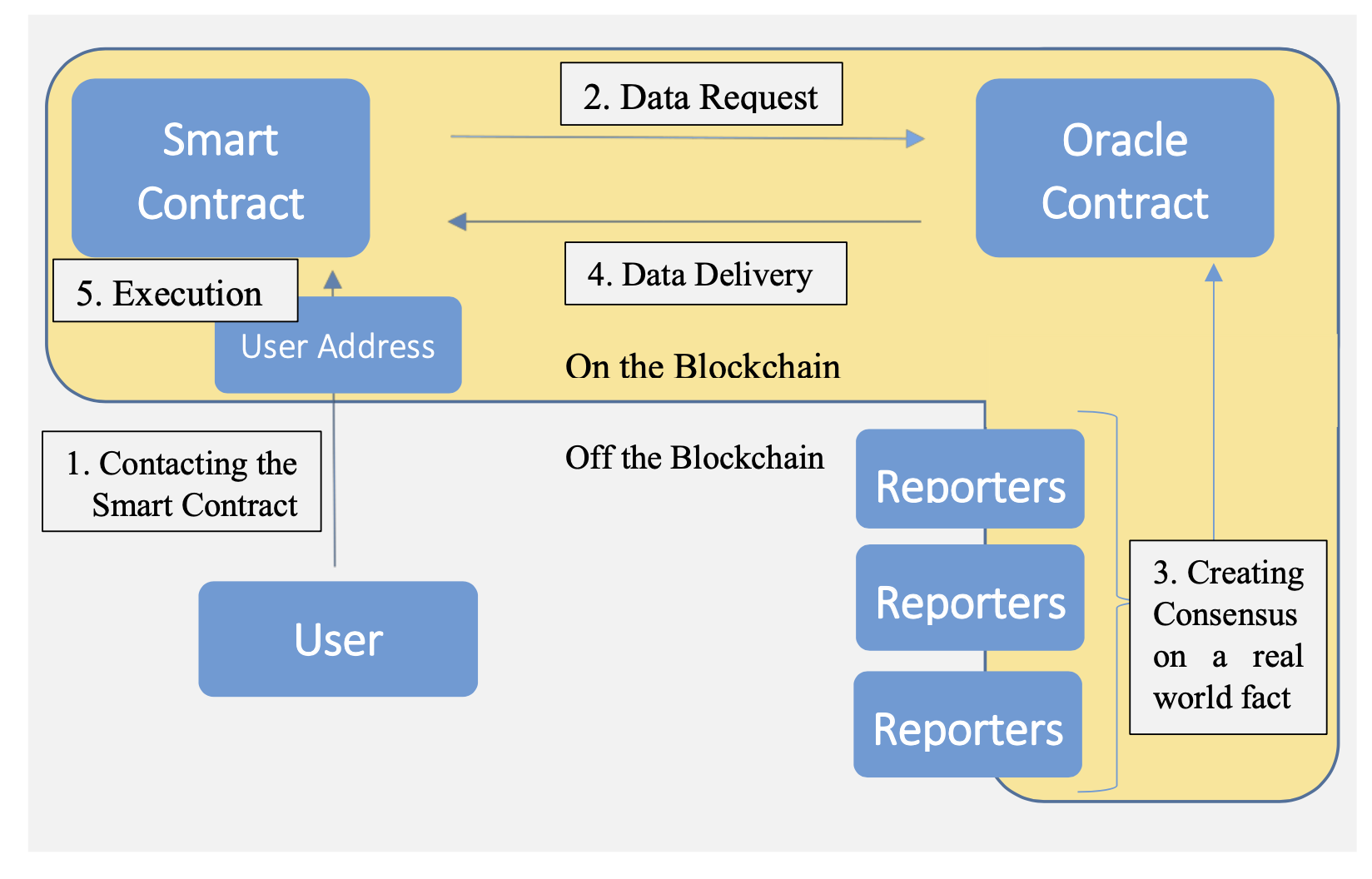}
        \caption{Human oracle structure}
\label{fig1}
\end{figure*}

\section{Oracle Design Patterns}

All types of oracles discussed provide a few key functions, by definition. These include the ability to:

\begin{itemize}
\item Collect data from an off-chain source.
\end{itemize}

\begin{itemize}
\item Transfer the data on-chain with a signed message.
\end{itemize}

\begin{itemize}
\item Make the data available by putting it in a smart contract's storage.
\end{itemize}

Once the data is available in a smart contract's storage, it can be accessed by other smart contracts via message calls that invoke a ``retrieve" function of the oracle's smart contract; it can also be accessed by Ethereum nodes or network-enabled clients directly by ``looking into" the oracle's storage.

The three main ways to set up an oracle can be categorized as (i) \emph{request-response}, (ii) \emph{publish-subscribe}, and (iii)\emph{ immediate-read}.

\subsection{Immediate-Read}

Starting with the simplest, immediate-read oracles are those that provide data that is only needed for an immediate decision, like ``What is the IP address for inrs.ca?" or ``Is this student over 25?" Those wishing to query this kind of data tend to do so on a ``just-in-time" basis; the lookup is done when the information is needed and possibly never again. Examples of such oracles include those that hold data about or issued by organizations, such as academic certificates, dial codes, institutional memberships, airport identifiers, self-sovereign IDs, etc.

This type of oracle stores data once in its contract storage, whence any other smart contract can look it up using a request call to the oracle contract. It may be updated. The data in the oracle's storage is also available for direct lookup by blockchain-enabled (i.e., Ethereum client-connected) applications without having to go through the palaver and incurring the gas costs of issuing a transaction. A shop wanting to check the age of a customer wishing to purchase alcohol could use an oracle in this way. 

This type of oracle is attractive to an organization or company that might otherwise have to run and maintain servers to answer such data requests. We note that the data stored by the oracle is likely not to be the raw data that the oracle is serving, e.g., for efficiency or privacy reasons. A university might set up an oracle for the certificates of academic achievement of past students. However, storing the full details of the certificates (which could run to pages of courses taken and grades achieved) would be excessive. Instead, a hash of the certificate is sufficient. Likewise, a government might wish to put citizen IDs onto the Ethereum platform, where clearly the details included need to be kept private. Again, hashing the data (more carefully, in Merkle trees with salts) and only storing the root hash in the smart contract's storage would be an efficient way to organize such a service.

\subsection{Publish-Subscribe}

The next setup is publish-subscribe, where an oracle that effectively provides a broadcast service for data that is expected to change (perhaps both regularly and frequently) is either polled by a smart contract on-chain, or watched by an off-chain daemon for updates. This category has a pattern similar to RSS feeds, WebSub, and the like, where the oracle is updated with new information and a flag signals that new data is available to those who consider themselves ``subscribed." Interested parties must either poll the oracle to check whether the latest information has changed, or listen for updates to oracle contracts and act when they occur. 

Examples include price feeds, weather information, economic or social statistics, traffic data, etc. Polling is very inefficient in the world of web servers, but not so in the peer-to-peer context of blockchain platforms: Ethereum clients have to keep up with all state changes, including changes to contract storage, so polling for data changes is a local call to a synced client. Ethereum event logs make it particularly easy for applications to look out for oracle updates, and so this pattern can in some ways even be considered a ``push" service. However, if the polling is done from a smart contract, which might be necessary for some decentralized applications (e.g., where activation incentives are not possible), then significant gas expenditure may be incurred.

\subsection{Request-Response}

The request-response category is the most complicated: this is where the data space is too huge to be stored in a smart contract and users are expected to only need a small part of the overall dataset at a time. It is also an applicable model for data provider businesses. In practical terms, such an oracle might be implemented as a system of on-chain smart contracts and off-chain infrastructure used to monitor requests and retrieve and return data. A request for data from a decentralized application would typically be an asynchronous process involving a number of steps. 

In this pattern, firstly, an externally owned account (EOA) transacts with a decentralized application, resulting in an interaction with a function defined in the oracle smart contract. This function initiates the request to the oracle, with the associated arguments detailing the data requested in addition to supplementary information that might include callback functions and scheduling parameters. Once this transaction has been validated, the oracle request can be observed as an EVM event emitted by the oracle contract, or as a state change; the arguments can be retrieved and used to perform the actual query of the off-chain data source. The oracle may also require payment for processing the request, gas payment for the callback, and permissions to access the requested data.

Finally, the resulting data is signed by the oracle owner, attesting to the validity of the data at a given time, and delivered in a transaction to the decentralized application that made the request, either directly or via the oracle contract. Depending on the scheduling parameters, the oracle may broadcast further transactions updating the data at regular intervals (e.g., end-of-day pricing information).

The steps for a request-response oracle may be summarized as follows:

\begin{itemize}
\item Receive a query from a DApp.
\end{itemize}

\begin{itemize}
\item Parse the query.
\end{itemize}

\begin{itemize}
\item Check that payment and data access permissions are provided.
\end{itemize}

\begin{itemize}
\item Retrieve relevant data from an off-chain source (and encrypt it if necessary).
\end{itemize}

\begin{itemize}
\item Sign the transaction(s) with the data included.
\end{itemize}

\begin{itemize}
\item Broadcast the transaction(s) to the network.
\end{itemize}

\begin{itemize}
\item Schedule any further necessary transactions, such as notifications, etc.
\end{itemize}

A range of other schemes are also possible; for example, data can be requested from and returned directly by an EOA, removing the need for an oracle smart contract. Similarly, the request and response could be made to and from an human, software, or hardware.

The request-response pattern described here is commonly seen in client-server architectures. While this is a useful messaging pattern that allows applications to have a two-way conversation, it is perhaps inappropriate under certain conditions. For example, a smart bond requiring an interest rate from an oracle might have to request the data on a daily basis under a request-response pattern in order to ensure the rate is always correct. Given that interest rates change infrequently, a publish-subscribe pattern may be more appropriate here especially when taking into consideration Ethereum's limited bandwidth.

Publish-subscribe is a pattern where publishers (in this context, oracles) do not send messages directly to receivers, but instead categorize published messages into distinct classes. Subscribers are able to express an interest in one or more classes and retrieve only those messages that are of interest. Under such a pattern, an oracle might write the interest rate to its own internal storage each time it changes. Multiple subscribed DApps can simply read it from the oracle contract, thereby reducing the impact on network bandwidth while minimizing storage costs.

In a broadcast or multicast pattern, an oracle would post all messages to a channel and subscribing contracts would listen to the channel under a variety of subscription modes. For example, an oracle might publish messages to a cryptocurrency exchange rate channel. A subscribing smart contract could request the full content of the channel if it required the time series for, e.g., a moving average calculation; another might require only the latest rate for a spot price calculation. A broadcast pattern is appropriate where the oracle does not need to know the identity of the subscribing contract.

\section{Use case of Oracle in DApps}

There exist a number of DApp in the market which use oracle as a mechanism for bridging the gap between the off-chain world and smart contracts. Some examples of data that might be provided by oracles include:

\begin{itemize}
\item Random numbers/entropy from physical sources such as quantum/thermal processes: e.g., to fairly select a winner in a lottery smart contract.
\end{itemize}

\begin{itemize}
\item Parametric triggers indexed to natural hazards: e.g., triggering of catastrophe bond smart contracts, such as Richter scale measurements for an earthquake bond.
\end{itemize}
\begin{itemize}
\item Exchange rate data: e.g., for accurate pegging of cryptocurrencies to fiat currency.
\end{itemize}
\begin{itemize}
\item Capital markets data: e.g., pricing baskets of tokenized assets/securities.
\end{itemize}
\begin{itemize}
\item Benchmark reference data: e.g., incorporating interest rates into smart financial derivatives.
\end{itemize}
\begin{itemize}
\item Static/pseudostatic data: security identifiers, country codes, currency codes, etc.
\end{itemize}
\begin{itemize}
\item Time and interval data: for event triggers grounded in precise time measurements.
\end{itemize}
\begin{itemize}
\item Weather data: e.g., insurance premium calculations based on weather forecasts.
\end{itemize}
\begin{itemize}
\item Political events: for prediction market resolution.
\end{itemize}
\begin{itemize}
\item Sporting events: for prediction market resolution and fantasy sports contracts.
\end{itemize}
\begin{itemize}
\item Geolocation data: e.g., as used in supply chain tracking.
\end{itemize}
\begin{itemize}
\item Damage verification: for insurance contracts.
\end{itemize}
\begin{itemize}
\item Events occurring on other blockchains: interoperability functions.
\end{itemize}
\begin{itemize}
\item Ether market price: e.g., for fiat gas price oracles.
\end{itemize}
\begin{itemize}
\item Flight statistics: e.g., as used by groups and clubs for flight ticket pooling.
\end{itemize}

\section{Conclusion}

As we have seen, oracles provide a crucial service to smart contracts: they bring external facts to contract execution. With that, of course, oracles also introduce a significant risk, if they are trusted sources and can be compromised, they can result in compromised execution of the smart contracts they feed.

Generally, when considering the use of an oracle, we have to be very careful about the trust model. If we assume the oracle can be trusted, we may be undermining the security of our smart contract by exposing it to potentially false inputs. That said, oracles can be very useful if the security assumptions are carefully considered.

Decentralized oracles can resolve some of these concerns and provide Ethereum smart contracts trust-less external data. We need to choose it carefully, then, we can start exploring the bridge between Ethereum and the ``real world" that oracles offer.


\begin{thebibliography}{1}


\bibitem{Event} Binance Academy, \textit{https://www.binance.vision/}.

\bibitem{Event} A. M. Antonopoulos and G. Wood, ``Mastering Ethereum: Building Smart Contracts and DApps,"\textit{O'Reilly Media Book (2019)}. 

\bibitem{Event} Provable documentation,  \textit{https://docs.provable.xyz/}.

\bibitem{Event} Provable White Paper.

\bibitem{Event} ChainLink documentation,  \textit{https://docs.chain.link/}.

\bibitem{Event} ChainLink White Paper.

\bibitem{Event} Ethereum Blog, \textit{https://blog.ethereum.org/}.

\bibitem{Event} Alexander Egberts, ``The Oracle Problem: An Analysis of how Blockchain Oracles Undermine The Advantages of Decentralized Ledger Systems," \textit{Master Thesis, EBS Universit{\"a}t f{\"u}r Wirtschaft und Recht}, December 2017.






\end{thebibliography}
\end{document}